\documentclass[12pt,a4paper]{article}
\pdfoutput=1
\usepackage{mathptmx}
\usepackage{amsmath}
\usepackage{amssymb}
\usepackage{amstext}
\usepackage{color}
\usepackage{graphicx}
\usepackage{url}
\usepackage{subfigure}
\usepackage{float}
\usepackage{sidecap}

\usepackage{ccaption}
\captionnamefont{\small\bfseries}
\captiontitlefont{\small\sloppy}
\usepackage{listings}
\newcommand{\bq}{\begin{equation}}
\newcommand{\eq}{\end{equation}}

\newcommand{\GBS}{\mbox{GB/s}}

\newcommand{\LUPS}{\mbox{LUP/s}}
\newcommand{\MLUPS}{\mbox{MLUP/s}}
\newcommand{\GLUPS}{\mbox{GLUP/s}}
\newcommand{\GHZ}{\mbox{GHz}}

\newcommand{\WF}{\mbox{W/F}}
\newcommand{\bytes}{\mbox{bytes}}

\newcommand{\KB}{\mbox{kB}}
\newcommand{\MB}{\mbox{MB}}

\newcommand{\mus}{\mbox{$\mu$s}}
\newcommand{\eos}{~.}

\begin{document}
\begin{center}\LARGE
Multicore-aware parallel temporal blocking of stencil codes for shared
and distributed memory\\[0.5cm]\large
Markus Wittmann, Georg Hager, Gerhard Wellein\\[1mm]\normalsize\itshape
Erlangen Regional Computing Center, 91058 Erlangen, Germany
\end{center}
\begin{abstract}
  New algorithms and optimization techniques are needed to balance the
  accelerating trend towards bandwidth-starved multicore chips.  It is
  well known that the performance of stencil codes can be improved by
  temporal blocking, lessening the pressure on the memory interface.
  We introduce a new pipelined approach that makes explicit use of
  shared caches in multicore environments and minimizes
  synchronization and boundary overhead.  For clusters of
  shared-memory nodes we demonstrate how temporal blocking can be
  employed successfully in a hybrid shared/distributed-memory 
  environment.
\end{abstract}

\section{Temporal blocking of stencil codes}

\subsection{Baseline and test bed}

The Jacobi algorithm is a simple method for solving boundary value problems. 
It serves here as a prototype for more advanced stencil-based methods like the
lattice-Boltzmann algorithm (LBM)\@. In three dimensions, one stencil (cell)
update is described by
\begin{equation}
 \label{eq:jacobi_kernel}
  B_{i,j,k} = \frac{1}{6}\left(A_{i-1,j,k} + A_{i+1,j,k} + A_{i,j-1,k}
             + A_{i,j+1,k} + A_{i,j,k-1} + A_{i,j,k+1}\right)\eos
\end{equation}
Successive sweeps over the computational domain are performed to reach
convergence, writing to grids $A$ and $B$ in turn. Counting the
usually required read for ownership (RFO) on a write miss, the kernel
has a naive code balance of $B_\mathrm{c} = 8/6\,\WF$ (double
precision words per flop)\@. However, the actual number of words
transferred over the slowest data path (the memory bus) can be reduced
to three per stencil update with a suitable spatial blocking
scheme. Furthermore, the RFO can be circumvented on current x86-based
processor architectures by employing non-temporal store instructions,
which bypass the cache hierarchy.  These and other well-known standard
optimizations like data alignment and SIMD vectorization have been
applied for our baseline version of the algorithm and are described in
the literature \cite{datta08,wellein09}\@. We use OpenMP-parallel C/C++ 
code whenever
possible, and revert to compiler intrinsics only if necessary,
e.g., if the compiler refuses SIMD vectorization because a loop
runs backward (as is the case with the compressed grid version
of pipelined temporal blocking, which will be described in 
Sect.~\ref{sec:ptb})\@.

The memory-bound performance of the baseline code on a given
architecture can be easily estimated by assuming that memory bandwidth
is the sole limiting factor, and that all other contributions can be
hidden behind it. This assumption is valid for current multicore
processors if all cores sharing a memory interface are used in a
parallel calculation, but may be false for single-threaded
code~\cite{thw09}\@. If the achievable STREAM COPY
bandwidth (using non-temporal stores) is $M_\mathrm{s}$, a ``perfect'' baseline Jacobi
code ($B_\mathrm{c}=0.33\,\WF$) should show a performance of
\bq
P_0 = \frac{M_\mathrm{s}}{16\,\bytes}\,\left[\LUPS\right]\eos
\eq
We use the ``lattice site updates per second'' (\LUPS)
metric here.
Unless otherwise noted, all benchmark tests were performed on a
cluster of dual-socket Intel Nehalem EP (Xeon 5550) compute nodes.
%(see Fig.~\ref{fig:nehalem}) 
The quad-core CPUs run at a clock speed of
2.66\,\GHZ\ and achieve a maximum STREAM COPY bandwidth of
18.5\,\GBS\ per socket, leading to an expectation of 2.3\,\GLUPS\ 
for a standard Jacobi algorithm in main memory. A shared 8\,\MB\ 
L3 cache is available to the four cores in a socket, while
L2 (256\,\KB) and L1D (32\,\KB) caches are exclusive to each 
core. Memory is physically distributed across the two sockets
but logically shared, forming a ccNUMA-type system.
The nodes are connected via with a fully non-blocking fat-tree
QDR-InfiniBand interconnect.

%\begin{SCfigure}[1.4]
%\includegraphics*[width=0.4\textwidth]{Images/Testbed_Nehalem_Layout}
%\caption{\label{fig:nehalem}Layout of an Intel Nehalem EP (Xeon 5550)
%  compute node.  The four cores on each socket share an 8\,\MB\ L3
%  cache, but also have separate L1D (32\,\KB) and L2 (256\,\KB)
%  caches. Memory is physically distributed but logically shared,
%  forming a ccNUMA-type system.}
%\end{SCfigure}

The idea behind temporal blocking is to perform multiple in-cache
updates on each grid cell before the result is evicted to memory,
thereby reducing effective code balance. Section~\ref{sec:ptb} 
will introduce a pipelined temporal blocking
scheme in a shared-memory parallel context, while
Sect.~\ref{sec:dmblock} describes how and under what conditions these
optimizations can be put to use in a hybrid
(shared/distributed-memory) code.

\subsection{Related work}

Improving the performance of stencil codes by temporal blocking is not
a new idea, and many publications have studied different methods in
depth~\cite{wonnacott00,jin01,kowarschik04,frigo05,kamil06}\@. However, the
explicit use of shared caches provided by modern multicore CPUs has
not yet been investigated to great detail. Ref.~\cite{wellein09}
describes a ``wavefront'' method similar to the one introduced here.
However, that work was motivated mainly by the architectural peculiarities
of multi-core CPUs, and does not elaborate on specific optimizations 
like avoiding boundary copies and optimizing thread synchronization. 
Our investigation is more general and explores a much larger parameter
space. Finally there is, to our
knowledge, as yet no systematic analysis of the prospects of temporal
blocking on hybrid architectures, notably clusters of shared-memory
compute nodes.

\subsection{Shared-memory pipelined temporal blocking on shared caches}\label{sec:ptb}

In contrast to previous approaches to cache reuse with stencil
algorithms, pipelined temporal
blocking makes explicit use of the cache topology on modern
processors, where certain cache levels are shared by groups of cores,
which we call \emph{cache groups}\@. Our Nehalem test system
%(Fig.~\ref{fig:nehalem}) 
is a typical example, where the four cores in
a socket share an 8\,\MB\ L3 cache.

\begin{figure}
\centering
\includegraphics[width=0.8\textwidth]{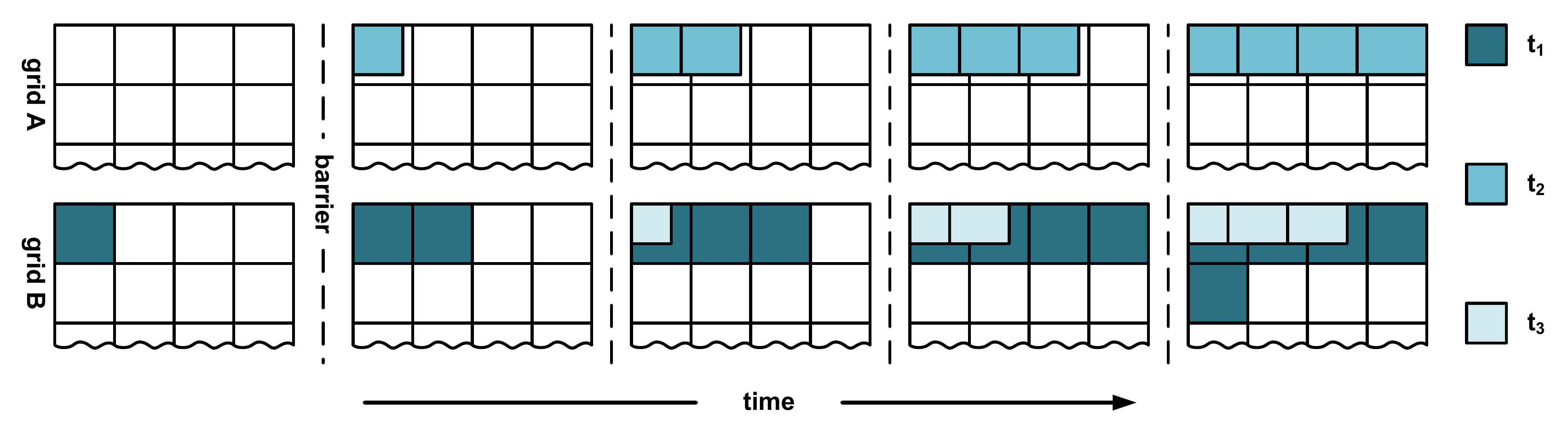}
\caption{\label{fig:pipelining}Temporal blocking by pipelining, shown
  here for three threads in two dimensions and separate grids $A$ and
  $B$\@.  All threads in a team update the most part of a block
  consecutively. Shifting the block by one cell in each direction
  after an update avoids extra boundary copies.  To avert race
  conditions, a global barrier is required after each block update.
  This scheme can be easily generalized to support multiple updates
  per thread. Possible optimizations include the use of a ``compressed
  grid'' update scheme, and relaxed synchronization (see text)\@.}
\end{figure}

Pipelined blocking splits the set of all available threads into
\emph{teams} of size $t$, where a team runs on cores sharing a cache.
It is possible that the size of a team is smaller than the whole cache
group, but this option will not be explored here. Each team has one
``front'' thread, which performs the first $T$ updates on a certain
grid block (see Fig.~\ref{fig:pipelining} for a visualization with
three threads and $T=1$)\@. Of course the block is loaded to cache in
this process, and if it is small enough, the remaining threads can
perform further updates ($T$ each) on it in turn before it gets
evicted to memory. All threads in the team can be kept busy by keeping
this pipeline running, with different blocks in different stages until
each block of the whole computational domain has been updated $t\cdot
T$ times, which completes a \emph{team sweep}\@. 
To avoid race conditions, the minimum distance between
neighboring threads is one block, but it may be larger. An estimate
for the maximum distance is given by the cache size divided by
$t$ times the size of one block. Due to the one-layer shift after each 
block update, the actual amount of cache needed is actually larger,
depending on the blocksize and the overall number of updates,
$t\cdot T$\@. In the simplest case, the distance is kept constant
by imposing a global barrier across all threads after each block
update. See below for ways to reduce synchronization overhead.

Pipelined blocking
has potential for automatic overlapping data transfer and calculation,
because the front thread continuously operates on new blocks, which
have to be fetched from memory. Compared to
the wavefront technique~\cite{wellein09}, it
does not incur extra work or boundary copies. 
The resulting kernel is fully
SIMD-vectorized, and the well-known ``compressed grid'' 
optimization can be applied here as well: During the first team
sweep, each result is written to a location shifted by the vector
(-1,-1,-1) relative to its original position. In order to avoid
complex address calculations, alternate team sweeps shift by
(-1,-1,-1) and (+1,+1,+1), respectively, requiring reverse loops
(running from large to small indices) on all even sweeps.  Since the
compiler refuses to properly SIMD-vectorize the inner loop in this
case, SSE intrinsics were used to get optimal code.  The use of
non-temporal stores is unnecessary and even counterproductive; after
the $t\cdot T$ updates in a team sweep, a block gets evicted to memory
automatically by the usual replacement mechanisms. The benefit of 
using ``compressed grid'' is that only one grid is necessary, saving
nearly half the memory and lessening the bandwidth requirements.

In typical shared-memory systems there is usually more than one
outer-level cache group, so more than one team must be kept running.
We choose to use those additional threads to perform further updates
on the blocks already handled by the ``front'' team. This enlarges the
whole update pipeline to $n\cdot t\cdot T$ stages, where $n$ is the
number of teams. Since different teams do not share any cache, blocks
updated by one team must be transferred to another cache when the next
team takes over. As will be shown in the results section, it makes
sense to enforce a larger distance between successive teams than
between neighboring threads inside a team. We call this extra distance
the \emph{team delay}, $d_\mathrm{t}$\@.

A problem with our method of consecutive thread teams is that every
thread updates every block, rendering ccNUMA placement optimizations
mostly useless. However, since the pressure on the memory interfaces
is greatly reduced by the temporal blocking, a round-robin page
placement strategy is adequate to achieve 
parallelism in memory access. The baseline Jacobi 
%and halo blocking
code employs standard first-touch page placement.

\subsubsection*{Relaxed synchronization}

It is well known that synchronizing a large group of threads running
on different cores can incur significant overhead. Depending on the
system topology (shared caches, sockets, ccNUMA locality domains), a
barrier may cost hundreds if not thousands of cycles even if
implemented efficiently~\cite{thw09}\@.
\begin{SCfigure}[0.63]
%\centering
\includegraphics[width=0.6\textwidth]{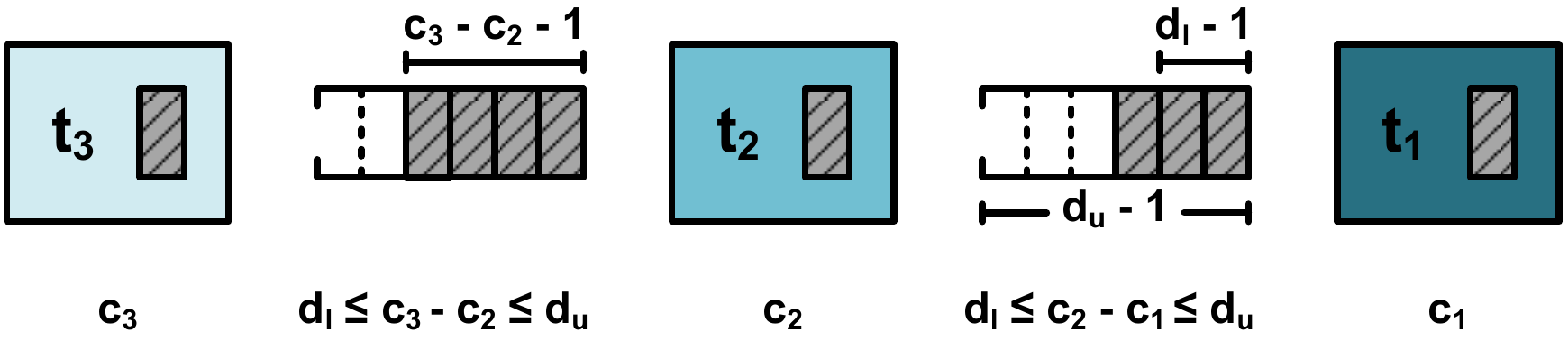}
\caption{\label{fig:relsync}Relaxed thread synchronization. A global
  barrier is avoided by observing ``soft'' lower and
  upper limits for the distance between consecutive threads.}
\end{SCfigure}
With the number of cores per shared-memory node increasing steadily
over time, alternatives should be used where appropriate. In our
pipelined temporal blocking scheme, the barrier can be removed
completely if the minimum and maximum thread distance and team delays
are still observed. To this end, each thread $t_i$ maintains a
(volatile) counter variable $c_i$, which is initialized to zero at the
start of each team sweep. It gets incremented whenever $t_i$ has
updated its current block. Each $c_i$ is located in a cache line of
its own to circumvent false sharing. The conditions to allow thread
$t_i$ to start updating the next block are then
\bq\label{eq:updcond}
c_{i-1}-c_i\geq d_\mathrm{l}\quad\wedge\quad c_i-c_{i+1}\leq d_\mathrm{u}\eos
\eq
The first condition averts data races, whereas the second maintains a
maximum distance between consecutive threads. The team delay is
trivially implemented by adding $d_\mathrm{t}$ to $d_\mathrm{l}$ on a
team's front thread and to $d_\mathrm{u}$ on its ``rear''
thread. Overall front and rear threads (i.e., the front thread of the
first and the rear thread of the last team) ignore the first and the
second condition, respectively.

In this scheme, only thread $t_i$ updates its own counter $c_i$; all
others read its updated value by means of the standard cache coherence
mechanisms.  Making the variables volatile prevents register
optimizations (using the OpenMP \verb.flush. directive instead would
be an alternative worth investigating)\@. The naive choice of
$d_\mathrm{l}=d_\mathrm{u}=1$ imposes a rigid lock-step between threads.
As will be shown in the results section, it is better to choose a different
value at least for $d_\mathrm{u}$, allowing for some 
``looseness'' in the pipeline.

\subsection{Single-cache diagnostic performance model}

One can try to establish an upper limit for the expected performance
gain through pipelined temporal blocking on a cache group. Following
the analysis in~\cite{wellein09}, we assume that the cache is large
enough to hold $(t-1)\cdot d_\mathrm{u}$ blocks, and that the
blocksize is chosen such that the shared cache must supply one
load and one store per stencil update only. Since all data that 
comes from memory must also be streamed through the shared cache,
the overall transfer time is the sum of in-memory and in-cache
contributions. The $t\cdot T$ block updates performed by a team
thus take
\bq\label{eq:ttime}
T_\mathrm{b} = \frac{16\,\bytes}{M_\mathrm{s,1}} 
	+ 2(t\cdot T-1) \cdot \frac{8\,\bytes}{M_\mathrm{c}} = 
	\frac{16\,\bytes}{M_\mathrm{s,1}}\left(
	 1 + (t\cdot T-1) \frac{M_\mathrm{s,1}}{M_\mathrm{c}} 
	\right)\eos
\eq
Here $M_\mathrm{s,1}$ is the memory bandwidth as obtained by a
\emph{single-threaded} STREAM COPY benchmark (a single stream is not
able to saturate the memory bus on most current multicore
processors), and $M_\mathrm{c}$ is the multi-threaded shared cache
bandwidth for STREAM COPY-like kernels. All upper cache levels are
assumed to be infinitely fast. The expected speedup compared to the
standard Jacobi algorithm is then
\bq
\frac{T_0}{T_\mathrm{b}} = 
%	\frac{t\cdot T\cdot 16\,\bytes/M_\mathrm{s}}
%	{16\,\bytes\left(
%	 1 + (t\cdot T-1) \frac{M_\mathrm{s,1}}{M_\mathrm{c}} 
%	\right)/M_\mathrm{s,1}} =
	\frac{M_\mathrm{s,1}}{M_\mathrm{s}}
	\frac{t\cdot T}{1+(t\cdot T-1)\frac{M_\mathrm{s,1}}{M_\mathrm{c}}}\eos
\eq
In the limit of very large $t\cdot T$, this ratio becomes 
$M_\mathrm{c}/M_\mathrm{s}$ as expected. The speedup increases if
$M_\mathrm{s,1}$ is close to $M_\mathrm{s}$, which is just another 
way of saying that the processor is ``bandwidth-starved'' when
using multiple cores to access memory. On the other hand, if the memory
bandwidth scales with core count, the factor of $t$ in the numerator
is canceled, making such an architecture a bad candidate for
temporal blocking. On the Nehalem system
we use, $M_\mathrm{s}/M_\mathrm{s,1}\approx 2$ and 
$M_\mathrm{c}/M_\mathrm{s,1}\approx 8$~\cite{thw09}, leading to
an expected speedup of $16T/(7+4T)$ at $t=4$, or 1.45 at $T=1$\@.
Although the model contains significant simplifications, our 
measurements on a single Nehalem socket match this prediction
almost exactly (see next section)\@. The maximum possible speedup 
on this CPU would be $M_\mathrm{c}/M_\mathrm{s}\approx 4$\@.

Note that our model assumes that code execution inside the caches and
also in memory is purely bandwidth-bound, and that the memory bus is
never idle, i.e., always saturated. If code execution decouples from
main memory bandwidth, or execution becomes limited by arithmetic
throughput, one cannot expect to get valid predictions.

\subsection{Socket and node results}\label{smresults}

We must stress that the parameter space for temporal blocking schemes,
and especially for pipelined blocking, is huge. The optimal
choices reported here have been obtained experimentally, with some
guidance from experience with older codes.

\begin{figure}
\includegraphics*[width=0.5\textwidth]{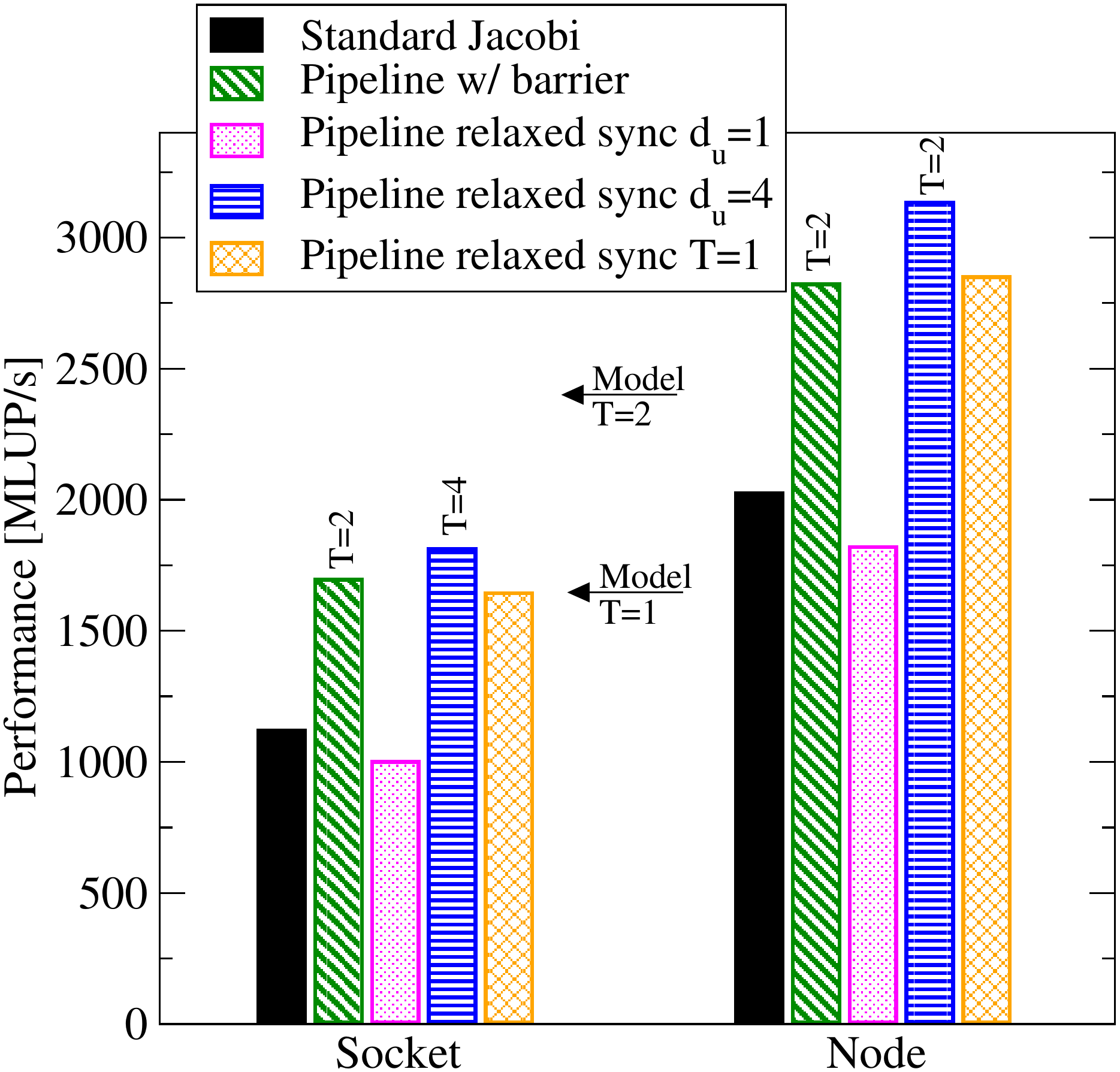}\hfill
\raisebox{3.6cm}{\includegraphics*[width=0.4\textwidth]{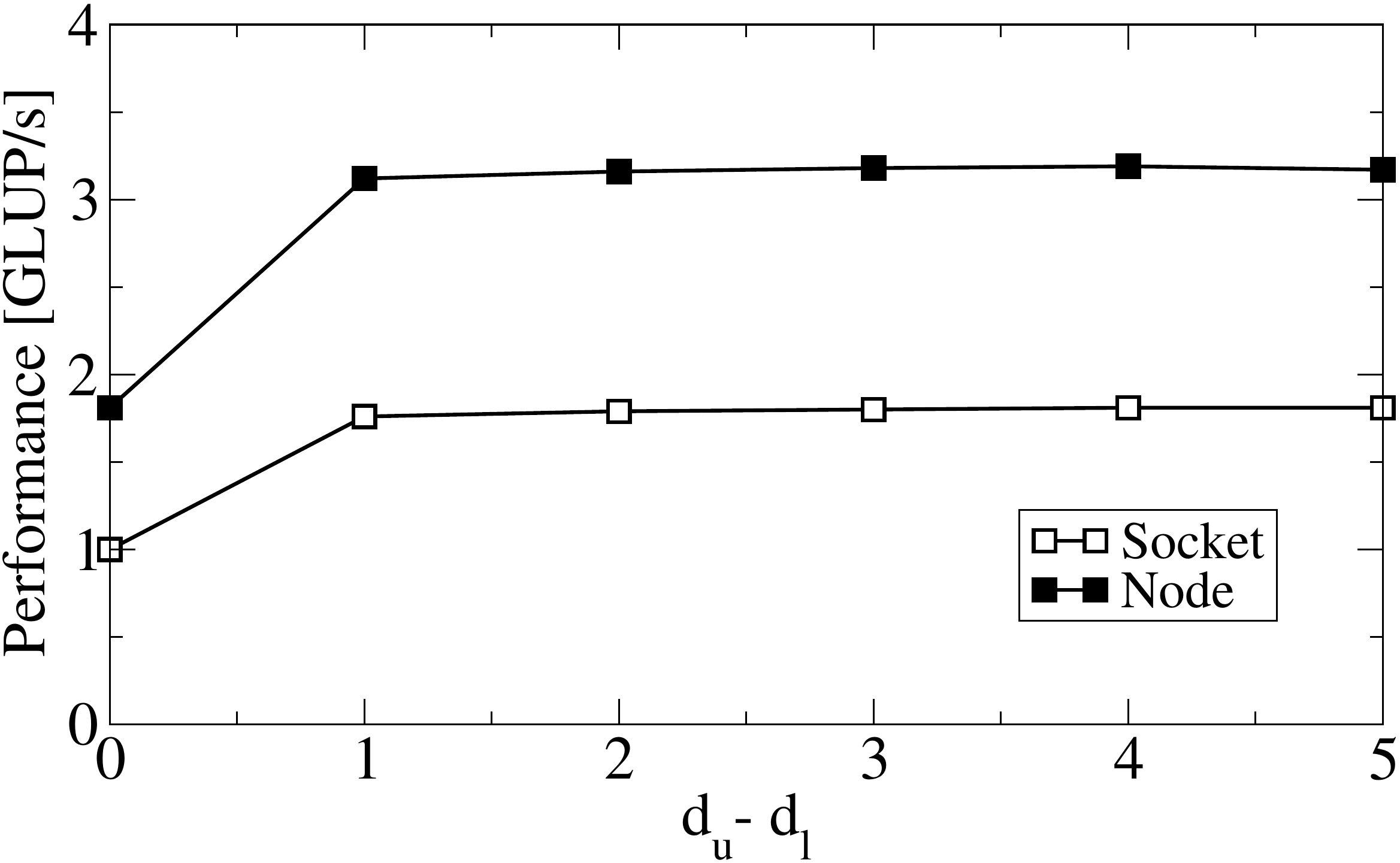}}~\hspace*{-0.45\textwidth}
\raisebox{3.6cm}{\begin{minipage}[t]{0.45\textwidth}\caption{\label{fig:node}
      Left: Single-socket and single-node results
	for pipelined blocking versus the standard version, 
        ($600^3$ grid)\@. $T=1$ and $T=2$
	predictions for one-socket performance are indicated.
	Optimal values for $T$ were determined empirically
	except for the $T=1$ data. Top: Influence of pipeline looseness.}
\end{minipage}}
\end{figure}
The left part of Fig.~\ref{fig:node} shows single-socket (one team) and node results
(two teams) on our test system for a fixed problem size of
$600^3$ (Intel compiler version 11.1.056 was used throughout)\@. 
The standard Jacobi data was  obtained with a blocksize
of roughly $600\times 20\times 20$ ($b_x\times b_y\times b_z$); it is
well known that due to the hardware prefetching mechanisms on current
x86 designs, a long inner loop (comparable to the page size) is
favorable~\cite{datta08}\@.  While the results are rather insensitive
to the blocksizes in $y$ and $z$ directions as long as the cache
size restrictions are observed, the inner loop length is also 
decisive for good performance on the temporally blocked versions.
Best performance is achieved around $b_x\approx 120$, with minor
variations depending on the number of time steps $T$ per thread.

%Halo blocking cannot even keep up with the standard Jacobi
%implementation. We attribute this failure to the large memory
%bandwidth of Nehalem in comparison to its cache bandwidth, and to the
%non-overlapping nature of the halo code. Experience with older
%architectures like Intel Core~2 has shown that halo blocking can be
%beneficial if the socket is strongly
%bandwidth-starved~\cite{wittmann09}\@.

In contrast, the pipelined blocking scheme can achieve speedups of up
to 50--60\,\% on one and two sockets. The relaxed synchronization
scheme is somewhat beneficial, and pays off most on two sockets where
the global barrier is much more costly than on the shared L3 cache of
a single processor. Nevertheless we expect it to be a vital
optimization on future many-core designs. The optimal number of
updates per thread and block, $T$, is usually 2 with some very minor
improvement at $T=4$\@. At $T=1$ the prediction from the diagnostic
performance model agrees perfectly with our measurements; however,
the model fails completely at larger $T$ for two reasons. First, code execution
has decoupled from main memory bandwidth already at $T=1$. With
four threads on the socket, a performance of 1600\,\MLUPS\ causes
a main memory bandwidth of about 6.4\,\GBS, which is just below
the single-thread STREAM COPY bandwidth of 10\,\GBS. Second, in-cache
performance for stencil codes is not dominated by bandwidth effects
alone, as has been shown in~\cite{thw09}\@.

The optimal range of values for $d_\mathrm{u}$, the upper limit for the 
distance of neighboring threads, was determined to be 1--4 with the 
block sizes chosen (see the right part of Fig.~\ref{fig:node})\@. 
This allows for sufficient looseness in the
pipeline without running the danger of blocks falling out of cache
before the team's rear thread has done its updates on them. Compared
to the ``lockstep'' case $d_\mathrm{l}=d_\mathrm{u}=1$, a performance
gain of about 80\,\% can be observed. Of course, 
$d_\mathrm{u}$ and the blocksize are strongly coupled, and larger
blocks would require smaller $d_\mathrm{u}$, but we could not find 
better combinations than the ones reported here.
A finite team delay $d_\mathrm{t}$
only has a very slight impact on this architecture (about 3\,\%
improvement for $d_\mathrm{t}=8$), and its influence will not
be studied further. 

Scalability of the pipelined code is not perfect across sockets, since
proper NUMA placement cannot be enforced, as described above. However,
distributed-memory parallelization (see next section) can employ one
process per socket, eliminating the need for first-touch parallel
placement.

\section{Distributed-memory parallelization}\label{sec:dmblock}

\subsection{Multi-halo exchange}

\begin{SCfigure}[2.2]
\includegraphics[width=0.3\textwidth]{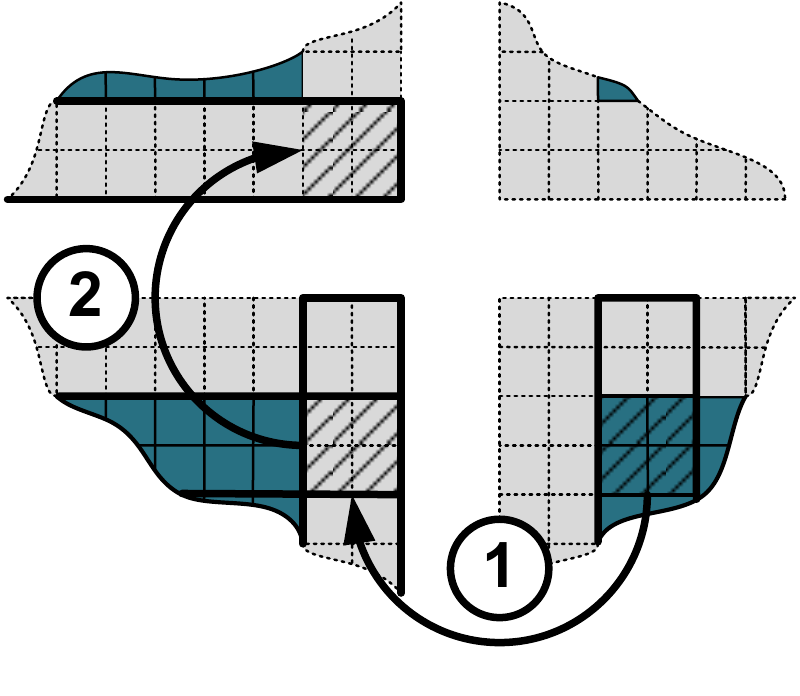}
\caption{\label{fig:dce}Multi-layer halo communication. Each
  halo is transmitted consecutively along the three coordinate
  directions, avoiding direct communication across edges and
  corners~\cite{ding01}.\bigskip}
\end{SCfigure}
The parallel temporal blocking schemes described above work on
multicore-based shared-memory architectures. Stencil algorithms are
usually straightforward to parallelize on dis\-tri\-bu\-ted-memo\-ry
systems using domain decomposition and halo layer exchange, and the
temporally blocked Jacobi code is no exception: The computational
domain is decomposed as usual, but instead of a single halo, $h$
layers must be exchanged after $h=n\cdot t\cdot T$ updates have been
performed per subdomain.  %Similar to the simple halo temporal blocking algorithm, 
Subdomains overlap by $h-1$ grid layers, and extra work is
involved on the boundaries because update number $s$ covers a domain
that is $h-s$ layers larger in each direction. The amount of data
communication per stencil update is roughly the same as for
no temporal blocking, except for edge and corner contributions,
which only become important on very small subdomains (see below)\@.
An important side effect of multi-layer halo exchange is that
communication takes place across subdomain edges and corners.
Latency-dominated small messages can be avoided by transmitting halos
consecutively along the three coordinate directions~\cite{ding01} (see
Fig.~\ref{fig:dce})\@.

The question arises whether the use of multi-layer halo exchange has
any significant impact on performance. Two factors lead to opposite
effects here: If the subdomain surface area is very small, aggregating
multiple messages into one may be beneficial; effective bandwidth
rises dramatically with growing message size in the latency-dominated
regime. On the other hand, the surface to volume ratio is large in
this limit, leading to a significant overhead from communication and
extra halo work.  
%A simple model will reveal whether the
%performance advantage from temporal blocking can be retained in the
%MPI-parallel case.
\begin{SCfigure}[0.63]%\centering
\includegraphics*[width=0.6\textwidth]{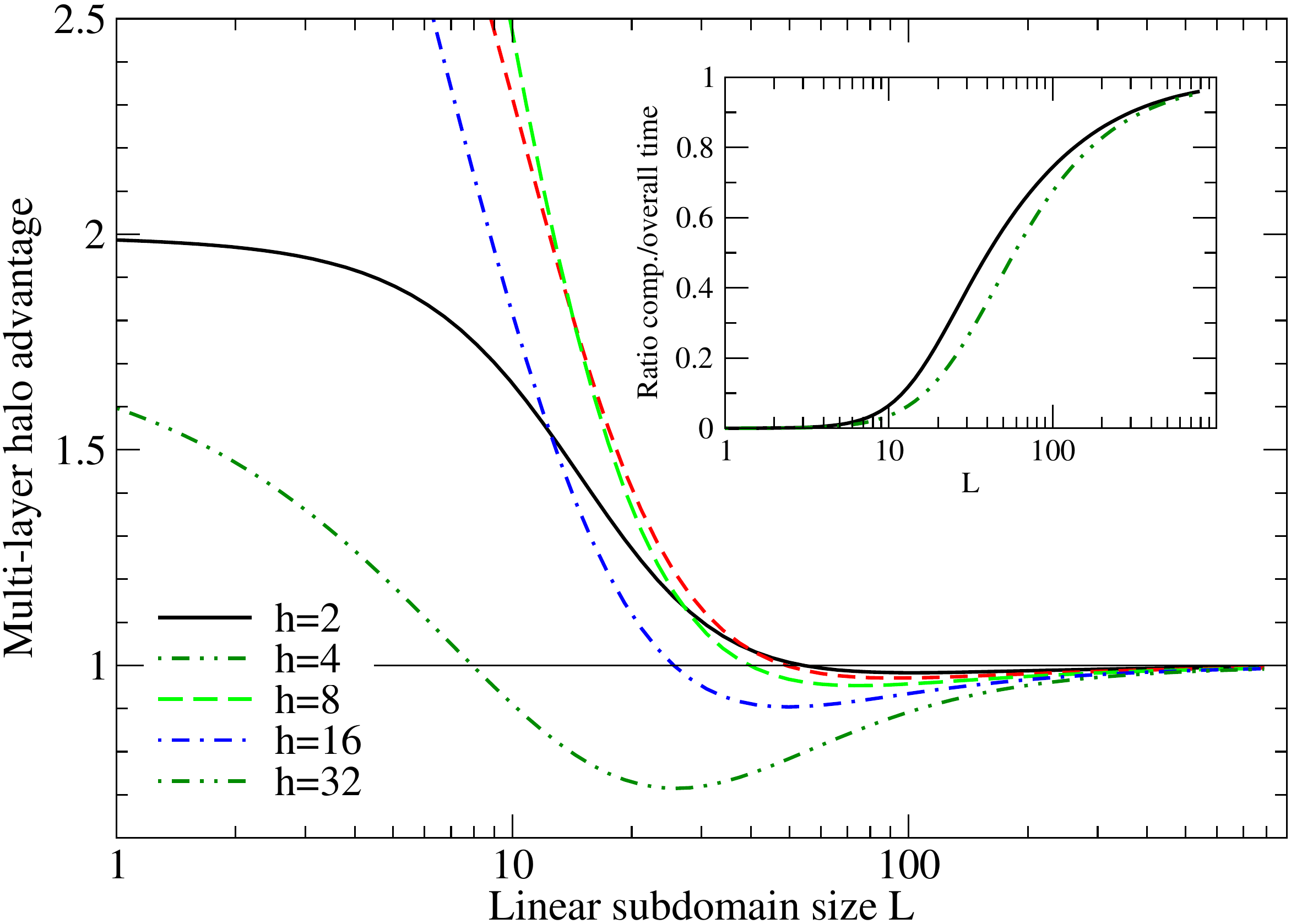}
\caption{\label{fig:mlh-adv}Theo\-re\-ti\-cal multi-layer halo advantage 
  versus linear subdomain size $L$ for different halo widths
  $h$\@. Parameters are set for a vector-mode hybrid Jacobi solver on
  a QDR-IB network and a per-node performance of 2000\,\MLUPS\ (see
  text for details)\@. Inset: Ratio of computation versus overall time
  (``computational efficiency'') for the corner cases $h=2$ and
  $h=32$\@.\bigskip}
\end{SCfigure}
%If $L$ is the linear subdomain size, $B$ is the asymptotic network
%bandwidth, $\ell$ is the network latency, and $P$ is the one-node
%performance, the time for $h$ bulk stencil updates $T_\mathrm{bulk}$,
The different contributions to execution time (``bulk'' and additional
``face'' stencil updates, and halo exchange) can be calculated,
assuming a simple latency/bandwidth model for network communication
and no overlap between calculation and data transfer. 

While only simple algebra is involved, the resulting expressions are
very complex, so we restrict ourselves to a graphical analysis. The
main panel of Fig.~\ref{fig:mlh-adv} shows the predicted ratio of
execution times between a standard one-layer halo version and
$h$-layer exchange for cubic subdomains of size $L^3$ and different
$h$\@. We have set the parameters for a QDR-InfiniBand network here,
with an asymptotic (large-message) unidirectional bandwidth of 3.2\,\GBS\ and a
latency of 1.8\,\mus\@. The single-node performance was assumed to be
2000\,\MLUPS, independent of $L$ (which only roughly holds in
practice)\@. As expected, multi-layer halos have no influence at large
subdomain sizes.  As the domain gets smaller ($20\lesssim L\lesssim
100$), extra halo work starts to degrade performance, but a relevant
impact can only be expected at $h\gtrsim 16$\@. At even smaller
$L\lesssim 20$, the positive effect of message aggregation
over-compensates the halo overhead, leading to substantial performance
gains. Although this looks like a good result, the ratio of
computation time versus overall execution time as shown in the inset
of Fig.~\ref{fig:mlh-adv} proves that the algorithm is strongly
communication-limited below $L\approx 100$, such that parallel
efficiency is very low. Any gain obtained by sophisticated
temporal blocking is squandered by communication overhead in this
limit.

Note that this simple model disregards some important effects like
switching of message protocols, overhead for copying to and from
message buffers, load imbalance, etc. Its purpose is to get a rough
idea about where to expect benefits from distributed-memory
parallelization of pipelined blocking with multi-layer halos.

\subsection{Distributed-memory results}\label{sec:dmresults}

Implementation of the hybrid MPI/OpenMP pipelined Jacobi code was
straightforward, with no explicit or implicit overlapping of
communication and computation. The MPI library used (Intel MPI
3.2.2.006) does not support asynchronous non-blocking transfers.
Profiling has shown that copying halo data from boundary cells to and
from intermediate message buffers causes about the same overhead as
the actual data transfer over the QDR-InfinBand interconnect.

\begin{SCfigure}[0.63]
\includegraphics*[width=0.6\textwidth]{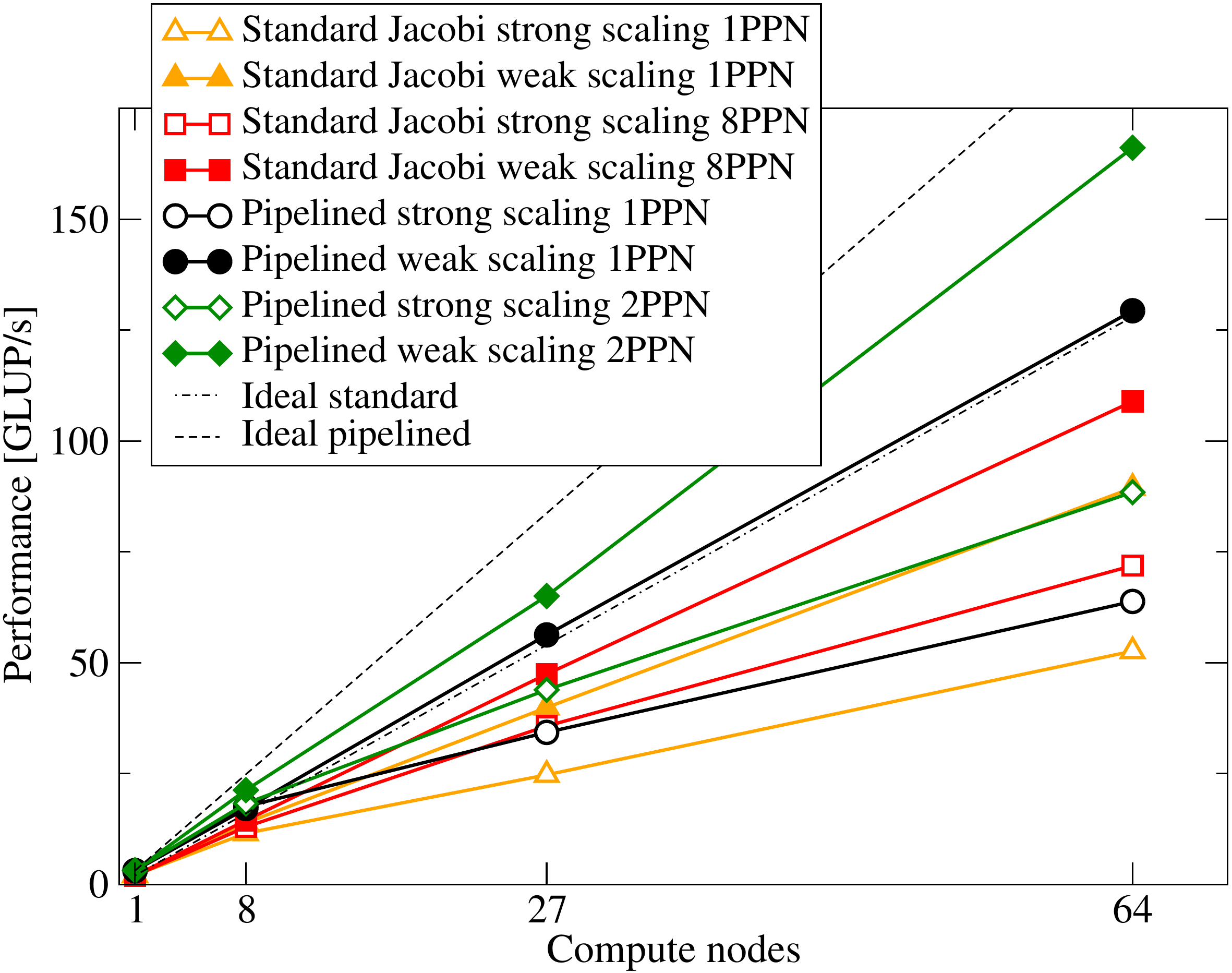}
\caption{\label{fig:dm}Distributed-memory parallel performance
	of the standard and the pipelined Jacobi solver with relaxed
	synchronization. Problem size was $600^3$ for strong scaling
	and $600^3$ per process for weak scaling. Ideal scaling
	behavior for both cases is indicated.\medskip}
\end{SCfigure}
Figure~\ref{fig:dm} shows performance data for strong and weak
scaling scenarios at problem sizes of $600^3$ overall 
and $600^3$ per node, respectively, on 1 to 64 nodes. The standard Jacobi 
code (triangles and squares) was run with one MPI process per
core (8PPN) and in ``hybrid vector'' mode (1PPN), %~\cite{rabe03},
the latter being clearly inferior. 

The results for the multi-halo pipelined code substantiate the
predictions from the model in the previous section: At large node
count and strong scaling, execution time is dominated by
communication. Hence, the benefit from temporal blocking cannot be
maintained in this case. On the other hand, much less performance
is lost with weak scaling.  Due to the ccNUMA page placement 
problems with pipelined blocking, using two processes per node
(2PPN), i.e., one per socket, yields a substantial improvement over
the 1PPN case. About 80\,\% of the pipelined blocking
speedup can be maintained for the distributed-memory parallel case.

\section{Summary and outlook}

We have demonstrated that multicore-aware pipelined temporal blocking
can lead to a substantial performance improvement for the Jacobi
algorithm on a current multicore architecture (Intel Nehalem
EP)\@. Substitution of a global barrier by relaxed synchronization
between neighboring threads adds to the benefits. At least for a
situation where the memory bus is still nearly saturated, a simple
bandwidth-based performance model can predict the expected speedup.
In comparison to earlier, more bandwidth-starved processor designs,
the potential gain on Nehalem is limited due to the small ratio
between cache and memory bandwidths, and the inability of a single
core to saturate the memory bus. However, future multicore processors
(just like the older Core 2 designs~\cite{wellein09,wittmann09})
can be expected to be less balanced, and thus profit more from
temporal blocking.
We have also shown, theoretically and in practice, under which
circumstances it is possible to port the temporal blocking speedup to
a distributed-memory parallel (hybrid) code. 
A hybrid, temporally blocked lattice Boltzmann flow solver
based on the principles presented in this work is under
development.

Further optimizations are possible: One main
drawback of our method is that all cores in a node form a single large
pipeline, inhibiting optimal ccNUMA placement. This could be corrected
by a domain decomposition strategy similar to the one demonstrated
in Ref.~\cite{wellein09}\@. Prospects for overlapping communication
and computation by functional decomposition among threads will be
investigated, but we expect substantial complexities in such an
implementation. We will also perform extensive benchmark tests on
other processor and system architectures, including massively
parallel systems like Cray XT and IBM Blue Gene.

\section*{Acknowledgments}

Lively discussions with Darren Kerbyson, Jan Treibig, Thomas Zeiser
and Johannes Habich are gratefully acknowledged. This work
was supported by BMBF under grant No.\ 01IH08003A (project SKALB)\@.

\end{document}